# PLASMA–WALL TRANSITION AND SHEATH FORMATION

### G. Manfredi and F. Valsaque

*Laboratoire de Physique des Milieux Ionisés et Applications*
*UMR 7040 du CNRS*
*Université Henri Poincaré, Nancy 1*
*F-54506 Vandœuvre-lès-Nancy*
*France*

### Abstract

A review of the theoretical and computational aspects of plasma–wall transition is presented, starting from the early contributions of Langmuir and Bohm. The conditions for the existence of plasma sheaths in front of a solid surface are established. Various regimes are analyzed — collisionless and collisional, fluid and kinetic, with and without ionization. The partition of the plasma–wall transition into distinct regions (Debye sheath, quasineutral presheath, magnetic presheath, etc…) is also discussed. A recently developed kinetic model is investigated in detail, both analytically and numerically, as a typical example of plasma–wall transition that is also relevant to magnetically confined plasmas encountered in thermonuclear fusion research. Applications to the theory of electrostatic probes are also highlighted.

### Résumé

Un examen des aspects théoriques et numériques de la transition plasma–paroi est présenté, à partir des contributions originales de Langmuir et Bohm. Les conditions d'existence de gaines devant une surface solide sont établies. Différents régimes sont analysés : avec et sans collisions ou ionisation, fluide et cinétique. La répartition de la transition plasma–paroi en plusieurs régions (gaine de Debye, prégaine quasineutre, prégaine magnétique, etc…) est aussi discutée. Un modèle cinétique récent est étudié en détail, analytiquement et numériquement, en tant qu'exemple typique de transition plasma–paroi d'intérêt pour la fusion thermonucléaire à confinement magnétique. Les applications à la théorie des sondes électrostatiques sont aussi soulignées.

**Mots clés** : Transition plasma–paroi, gaines, théorie des sondes, fusion nucléaire.

## I. INTRODUCTION

The interaction of a plasma with a surface is one of the oldest problems in plasma physics. This is hardly surprising, as any plasma created in the laboratory needs to be confined by a material vessel. Besides, a large number of diagnostics are obtained with probes inserted into the plasma, thus exposing some solid surface to the charged particles. The fact that plasma–surface interaction is still a lively subject of research seventy years after the publication of Tonks and Langmuir's seminal paper [1], is a clear sign that the problem is not an easy one. Indeed, the first (solid) and fourth (plasma) states of matter do not co-exist peacefully. Energetic plasma particles strike the solid surface ('sputtering'), in a process that can end up in significant erosion of the surface. Meanwhile, sputtered atoms can leave the surface and enter the plasma, leading to its contamination.

Perhaps the most distinctive character of plasma–wall interaction is that solid surfaces act as sinks and sources for the plasma. When an ion hits the surface, it is usually retained on it for a time sufficiently long to recombine with the electrons on the surface. The atoms thus formed are usually weakly bounded to the surface, and are re-emitted as neutrals into the plasma. Subsequently, these neutrals can be re-ionized, generally by electron impact. Depending on the plasma parameters, re-ionization can occur close to the surface or further into the bulk plasma. Besides, electrons can also be emitted by the surface by impact of other particles (including other electrons). This process of recombination/emission/ionization can lead to a steady-state regime named *recycling*. The most common example of recycling is provided by the familiar 'neon' light (indeed, a low-pressure gas discharge tube), in which ions are produced by electron impact in the cylindrical volume, then travel to the wall where they are recycled. Energy is lost to the wall as heat (except a small part ending up in visible light – the intended effect!), and must be provided continuously via an electric current that causes ohmic heating of the electrons [2].

A second, and most important, effect caused by the presence of a solid surface is the formation of plasma *sheaths*. Indeed, ions and electrons hit the surface at very different rates, roughly proportional to their thermal speeds, which scale as the square root of the ion-to-electron mass ratio (for equal temperatures, the electron thermal speed is about forty times that of hydrogen ions). If the surface is an insulator, or kept electrically isolated, a net charge develops on it and perturbs the ambient electric field, temperature and plasma flow, as well as other crucial parameters. This perturbation of the plasma, characterized by the presence of a distinct space charge, is called the *Debye sheath*. The Debye sheath (DS in the following) is crucial in mediating the transition from the unperturbed plasma to the wall, and its physics is fairly well understood. However, the DS cannot be directly connected to the unperturbed plasma. It must be preceded by a quasineutral region, the *presheath*, which is dominated by collisions and/or ionization, whereas the DS is essentially collisionless (in more complex cases, the presheath may also be determined by geometrical or magnetic effects). The role of the presheath is to accelerate the ions to a critical velocity (typically, the sound speed) at the entrance of the DS. Such a condition, named *Bohm's criterion*, is a cornerstone of plasma–wall interaction research, and has been the source of innumerable debates over the years [3].

Potential applications of plasma–wall interaction are ubiquitous, but a significant number of them are related to magnetically confined plasmas for nuclear fusion research. The plasma is confined by strong magnetic fields in a toroidal chamber (*tokamak*); however, radial transport towards the walls cannot be completely suppressed, and some charged particles reach the edge of the chamber. In order to optimize the interactions of the plasma with the external walls, some special configurations have been implemented, known as *limiters* and *divertors*. The physics of these devices is rather complex and beyond the scope of the present paper – a recent and comprehensive review is given in [2]. However, simple one-dimensional models, such as those investigated here, can be surprisingly accurate in reproducing the main features of the sheaths.

Another domain of applications of plasma–wall interactions in magnetic fusion plasmas comes from probe theory. Probes are routinely used for tokamak edge measurements, though their results are notoriously difficult to interpret, because the very presence of the probe can perturb the ambient plasma by creating a sheath. It is therefore of paramount importance to assess the properties of the sheath in order to relate the quantities measured by the probe to those of the unperturbed plasma.

The present paper is concerned with the properties of plasma–wall interaction from the viewpoint of plasma dynamics, and particularly sheath formation. The complex atomic processes involved at the plasma–surface boundary, although important in practice, will not be considered here, and only simple models of ionization and collisions will be adopted. The basic properties of plasma sheaths, and a review of some of the pertinent models, are presented in Sec. II. Section III illustrates the rich physics of plasma–wall transition by using numerical simulations of a relatively simple one-dimensional (1D) kinetic model, which has been recently applied to model measurements obtained from retarding-field analyzer probes. Conclusions are reported in Sec. IV.

## II. PLASMA SHEATHS

### II.1 Bohm's criterion and the Debye sheath

In order to fix the ideas, let us consider the simplest possible model of a sheath [4]: the ions obey a system of fluid equations with zero pressure, whereas the electrons are at thermal equilibrium with temperature $T_e$. For a one-dimensional system, the continuity and momentum equations for ions of mass $m_i$ and charge $e$ read as

$$\frac{\partial n_i}{\partial t} + \frac{\partial (n_i u_i)}{\partial x} = 0 \tag{1}$$

$$\frac{\partial u_i}{\partial t} + u_i \frac{\partial u_i}{\partial x} = -\frac{e}{m_i} \frac{\partial \Phi}{\partial x}, \tag{2}$$

where $n_i$ and $u_i$ are the ion density and average velocity, and $\Phi$ is the electrostatic potential, satisfying Poisson's equation

$$\frac{\partial^2 \Phi}{\partial x^2} = -\frac{e}{\varepsilon_0} (n_i - n_e) \tag{3}$$

The system is closed by specifying that the electrons follow the Boltzmann relation

$$n_e(x,t) = n_0 \exp\left(\frac{e\Phi}{k_B T_e}\right), \tag{4}$$

where $n_0$ is the equilibrium density and $k_B$ is Boltzmann's constant.

By considering stationary states, Eqs. (1)–(2) can be easily integrated between $x = 0$ (position of the wall, see Fig. 1) and $x = -\infty$ (unperturbed plasma). In the unperturbed plasma, we take the equilibrium values: $n = n_0$, $u = u_0$ and $\Phi = 0$. Solving Eqs. (1)–(2) for the ion density yields

$$\frac{n_i}{n_0} = \left(1 - \frac{2e\Phi}{m_i u_0^2}\right)^{-1/2}. \tag{5}$$

Substituting Eqs. (4)–(5) into Poisson's equation (3) yields a nonlinear differential equation for the electrostatic potential, which cannot be completely integrated analytically. However, one can linearize the expressions for the electron and ion densities by making the additional assumption that the potential energy is small. In this case, the linearized Poisson equation becomes

$$\frac{d^2\Phi}{dx^2} = \frac{1}{\lambda_D^2}\left(1 - \frac{k_B T_e}{m_i u_0^2}\right)\Phi, \qquad (6)$$

where $\lambda_D = \sqrt{e^2 n_0 / \varepsilon_0 k_B T_e}$ is the Debye length. In order for the potential to be a monotonic function, one needs to satisfy

$$u_0 > \sqrt{\frac{k_B T_e}{m_i}} \equiv c_s. \qquad (7)$$

The previous relation is known under the name of *Bohm's criterion* [5], and specifies that the ion mean velocity at the entrance of the sheath must be larger than the acoustic velocity $c_s$ (here computed with zero ion temperature, because of the simplified model adopted at this stage). Further considering the linearised Eq. (6), we see that the solution is an exponential function with characteristic decaying length close to the Debye length (as long as the electron temperature is not too high). Therefore, the potential will be appreciably different from zero in a region of thickness $\lambda_D$ in front of the wall (Fig. 1): this region is named the *Debye sheath*.

The typical structure of the Debye sheath is shown in Fig. 1, and confirms the results of the linear analysis. Note that the potential is everywhere negative, and that the ion density always exceeds the electron density. The latter statement is actually another form of Bohm's criterion, as can be proven by plotting both densities, given by Eqs. (4)–(5), as a function of the potential.

Physically, the existence of the Debye sheath can be understood as follows. In a plasma, the electrons are generally more mobile than the ions, as their thermal speed $v_T = \sqrt{k_B T / m}$ is much larger. Therefore, they are lost more quickly to the wall, and leave a net positive charge in the plasma, which must then be at a higher potential with respect to the wall. As we have chosen (arbitrarily) the plasma potential to be equal to zero, it follows that the sheath potential must be negative. This negative wall potential tends to reduce the electron flux and to increase the ion flux to the wall. At equilibrium, the two fluxes exactly balance one another, so that the net charge current is zero. In other words, the wall potential adjusts itself self-consistently to a negative value, such that exactly as many ions and electrons hit the wall per unit time. Naturally, because of the screening effect, the potential cannot extend indefinitely into the plasma, but is confined to a thin layer of the order of the Debye length (note that $\lambda_D$ is a few micrometers for a fusion reactor and few millimeters for a typical glow discharge).

The value of the potential at the wall can be estimated by noticing that the electron velocity distribution will be approximately half-Maxwellian (because the electrons are at thermal equilibrium and the wall is absorbing). Therefore the electron current is mainly determined by the thermal velocity

$$J_e^{\text{wall}} = n_e v_{Te} = n_0 v_{Te} \exp(e\Phi_{\text{wall}} / k_B T_e). \qquad (8)$$

The ion current is constant within the DS and equal to $J_i^{\text{wall}} = n_0 u_0 \cong n_0 c_s$, where we have assumed (following Bohm's criterion) that the ion speed is of the order of the sound speed at the DS entrance. At the wall, the currents must be equal, which allows us to obtain the potential

$$\frac{e\Phi_{\text{wall}}}{k_B T_e} = \frac{1}{2} \ln \frac{m_e}{m_i}. \tag{9}$$

The previous expression shows that the value of the wall potential is a few times the electron thermal energy (3.76 for hydrogen ions and 4.10 for deuterium). This is in good agreement with more sophisticated estimations.

Finally, we stress that the qualitative structure of the Debye sheath discussed above is rather general and does not depend on the particular (and very simplified) model adopted here.

## II.2 The presheath

We have established so far that an electrically charged layer (Debye sheath) is formed when a plasma is in contact with a solid surface and that, within the DS, the ions must travel at a speed at least equal to the acoustic velocity. However, at some distance from the surface, the core plasma must be at rest. There must therefore exist an intermediate region in between the core and the DS where the plasma is accelerated up to the acoustic velocity: this region is termed the *presheath*. It will appear in the forthcoming analysis that the presheath is *quasineutral* (i.e. $n_e = n_i$), although an electric field is present, whose role is precisely to accelerate the plasma to $c_s$. Collisions (ion-ion or ion-neutral) and/or ionization play a crucial role in the presheath (whereas they can be neglected in the DS), so that the typical extension of the presheath is given by the collision or ionization length $L_c$, rather than the Debye length $\lambda_D$. Quasineutrality arises naturally in the limit $\lambda_D \ll L_c$, which is generally true for fusion plasmas. By rewriting Poisson's equation (3) in dimensionless units (the potential is normalized to $k_B T_e/e$, densities to $n_0$ and space to $L_c$), we obtain

$$\frac{\lambda_D^2}{L_c^2} \frac{d^2 \Phi}{dx^2} = n_e - n_i. \tag{10}$$

Clearly, in the limit of vanishing Debye length, one obtains that $n_i = n_e$. The entrance of the DS can be defined as the point at which quasineutrality breaks down and a space charge appears. Such a point can be defined rigorously only in the limit $\lambda_D \to 0$, and corresponds to a mathematical singularity in the equations (see end of this section). For small, but finite, values of the Debye length, the transition between the presheath and the DS is smooth, and the DS entrance cannot be defined without some ambiguity. Also notice that for $\lambda_D \to 0$ the DS collapses to a single point, which coincides with the DS entrance.

We shall now make the previous arguments more quantitative. Let us express the ion density through the definition of the ion current: $n_i = J_i / u_i$. Taking the logarithmic derivative and rearranging some terms, yields [3]

$$\frac{1}{J_i}\frac{dJ_i}{dx} = \frac{1}{m_i u_i^2}\frac{d}{dx}\left(\frac{m_i u_i^2}{2}\right) + \frac{d}{dx}\left(\frac{e\Phi}{k_B T_e}\right). \tag{11}$$

In the presheath, where Bohm's criterion is *not* satisfied, we have that $u_i \leq c_s$ Therefore we can write

$$\frac{1}{J_i}\frac{dJ_i}{dx} \geq \frac{1}{k_B T_e}\frac{d}{dx}\left(\frac{m_i u_i^2}{2}\right) + \frac{d}{dx}\left(\frac{e\Phi}{k_B T_e}\right) \equiv \frac{d}{dx}\left(\frac{E}{k_B T_e}\right), \tag{12}$$

where we have defined the total energy $E$. Notice that, inside the DS, both energy and current are conserved, so that both sides of Eq. (11) are equal to zero. In order to satisfy Eq. (12), some conditions must be verified, which correspond to different physical situations:

a) Energy is lost ($dE / dx < 0$), but current is conserved ($dJ_i / dx = 0$). This corresponds to a case with friction (collisions), but no ionization.
b) Energy is conserved ($dE / dx = 0$), but current increases towards the wall ($dJ_i / dx > 0$). This corresponds to a case with ionization, but no ion collisions.
c) Energy is lost ($dE / dx < 0$) and current increases towards the wall ($dJ_i / dx > 0$). Both ionization and collisions are present.

Notice that the current increase, apart from ionization, can also arise from curved (e.g. cylindrical or spherical) geometry. Indeed, the ion continuity equation at steady state reads

$$\nabla \cdot J_i \equiv \frac{1}{x^\beta}\frac{d}{dx}\left(x^\beta J_i\right) = 0, \tag{13}$$

where $\beta = 0, 1, 2$ respectively for plane, cylindrical and spherical 1D geometry. Therefore, for $\beta \geq 1$, the current increases towards the wall because the elemental area is not constant in curved geometry.

In order to study the properties of the presheath in some detail, we slightly generalize the model previously used, Eqs. (1)–(2), to include the effect of ionization, collisions and geometry. At steady state, the ion continuity and momentum equations read

$$\frac{d}{dx}(n_i u_i) = S(x) \tag{14}$$

$$u_i \frac{d u_i}{dx} = -\frac{e}{m_i}\frac{d\Phi}{dx} - W u_i, \tag{15}$$

where $S(x)$ is a source, representing ionization and/or geometrical effects, and $W$ is the collision frequency, which we take to be constant for simplicity. These equations are completed by the quasineutrality condition (obtained by equating the ion and electron densities, and using Boltzmann's relation for the latter)

$$\frac{e\Phi}{k_B T_e} = \ln\frac{n_i}{n_0}. \tag{16}$$

Some algebra on Eqs. (14)–(16) enables us to rewrite them as

$$(u_i^2 - c_s^2)\frac{du_i}{dx} = -Wu_i^2 - c_s^2 \frac{S}{n_i}. \qquad (17)$$

Equation (17) presents a singularity for $u_i = c_s$. This singularity corresponds to the entrance of the DS, and reveals that the above model is no more appropriate when the ion speed reaches the acoustic velocity. Indeed, within the Debye sheath, we should have employed the complete Poisson's equation instead of assuming quasineutrality; by doing so the singularity would have been removed, and the complete model would describe a smooth transition between the presheath and the DS.

Let us first consider the case $S = 0$ and $W \neq 0$. Equation (17) can be integrated with boundary condition $u_i(x = 0) = c_s$, where $x = 0$ represents the entrance of the DS, yielding

$$u_i + \frac{c_s^2}{u_i} + Wx - 2c_s = 0. \qquad (18)$$

This is a quadratic equation for $u_i$, whose solution (using the rescaled variable $y = Wx/2$) is: $u_i = c_s - y - \sqrt{y^2 - 2yc_s}$. In the core plasma ($x \to -\infty$), the ion mean velocity $u_i$ tends to zero, in agreement with the reasonable assumption that the plasma is at rest far from the wall. At the wall ($x = 0$), $u_i$ is equal to the acoustic velocity, as required by the boundary condition, but its derivative becomes infinite, signaling the presence of a singularity. Further, notice that the presheath described by Eq. (18) extends over an infinite distance. This is because we have neglected the source term (i.e. ionization or geometrical effects) in our treatment. Including these effects shows that the presheath can have a finite thickness.

Turning to the case $W = 0$ and $S \neq 0$, an interesting form of the source function is the one investigated in [6]

$$S(\Phi) = \Omega\, n_0 \exp(\gamma\, e\Phi / k_B T_e) = \Omega\, n_0 \left(\frac{n_e}{n_0}\right)^\gamma \qquad (19)$$

where $\Omega$ is a typical ionization frequency. The relevant cases are those with $\gamma = 0$, 1 and 2, corresponding respectively to ion generation that is spatially uniform, proportional to the electron density, or proportional to the square of the electron density. Physically, such three cases represent ionization by a uniform electron beam, ionization by electron-neutral collisions, and ionization by a two-stage process involving excited atoms. Kino and Shaw [6] produced analytical solutions for all the above values of $\gamma$ in planar geometry, and for $\gamma = 0$ in cylindrical geometry [The case $\gamma = 1$ is particularly simple algebraically, as the last term in Eq. (17) reduces to a constant, so that Eq. (17) can be integrated immediately, yielding a cubic equation for $u_i$]. Other cases were solved numerically. Further, Kino and Shaw extended the above model to include the effect of ion pressure, and considered a two-dimensional case, both rectangular and cylindrical. All these results were obtained in the collisionless case, i.e. $W = 0$. More recently, Franklin and Snell [7] have analyzed the presheath–Debye sheath structure with fluid equations for both species of charged particles (i.e. without the assumption of Boltzmann distributed electrons). They discussed the conditions under which the Bohm criterion remains valid in the presence of collisions and ionization.

The typical structure of the presheath is plotted in Fig. 2, for a case with zero Debye length (thus, the DS entrance and the wall collapse to a single point). The ion velocity is accelerated from zero in the unperturbed plasmas to $c_s$ at the wall, while the ion and electron densities decrease. Note that the DS entrance is a singular point in the quasineutral case, where both the potential and the ion mean velocity have an infinite derivative. Figure 3 shows the overall sheath (presheath plus DS) for a case of small, but finite, Debye length, which removes the singularity. The DS is the region where quasineutrality breaks down. Notice the sharp increase in the potential gradient at the entrance of the DS.

## II.3 The magnetic presheath

In this section, we shall briefly describe the effect of a magnetic field on the sheath, which has been neglected so far. First, we notice that if the magnetic field is normal to the wall, then it has no effect on the structure of the sheath, at least for collisionless plasmas. Therefore, some of the results presented in the previous sections also apply to the case of a normal magnetic field.

In the simplest situation, the magnetic field is uniform and time-independent, and makes an angle $\alpha$ with the wall. This case was studied by Chodura [8] using a collisionless kinetic model with no ionization, for *both* ions and electrons, i.e., without using the Boltzmann approximation for the latter (see next section for some details on kinetic modeling). The full Poisson equation was included, thus permitting the formation of a Debye sheath. Chodura performed numerical computations with a particle-in-cell (PIC) codes for different values of the angle of incidence and other crucial parameters (e.g., mass and temperature ratios). The pertinent phase space for this physical situation is at least four-dimensional, including one spatial coordinate (the distance from the wall) and three velocity components.

The numerical results showed that, even in the absence of collisions and ionization, the plasma–wall transition presents a double structure: the usual DS in the immediate vicinity of the wall, and the *magnetic presheath* (MPS). The latter is a quasineutral region that extends over several Larmor radii, defined as $\rho_i = v_{Ti}/\Omega_i$, where $\Omega_i = eB/m_i$ is the ion cyclotron frequency (frequency of gyration of a charged particle around the magnetic field lines). Notice that, for magnetically confined fusion plasmas, the Larmor radius is much larger than the Debye length, so that the MPS extends over a larger region compared to the DS. The numerical results also showed that the potential at the wall is roughly independent of the angle of incidence. Chodura's results were recently confirmed and extended by simulations performed with a more accurate Eulerian Vlasov code (although Boltzmann electrons were assumed) [9]. Chodura also proposed a condition for the magnetic presheath, namely that the average speed *parallel to B* at the entrance of the MPS must exceed the acoustic velocity: $v_{//} \geq c_s$. This must be compared with Bohm's criterion, which states that the average speed *normal to the wall* must exceed $c_s$ at the entrance of the DS.

Some further analytical results on the plasma–wall transition in an oblique magnetic field can be found in references [10–14].

## II.4 Kinetic modeling

The plasma models presented in Secs. II.1 and II.2 are all of the fluid type. However, weakly collisional plasmas should in principle be described by a kinetic equation of the Boltzmann type. Kinetic equations describe the evolution of a particle distribution function, which is a function of the phase space variables, $f(x,v)$. The quantity $fdxdv$ represents the number of particles contained in the elementary phase space volume. The 1D Boltzmann equation for the ions can be written as

$$\frac{\partial f}{\partial t} + v \frac{\partial f}{\partial x} - \frac{e}{m_i} \frac{\partial \Phi}{\partial x} \frac{\partial f}{\partial v} = \overline{S}(f), \qquad (20)$$

where $\overline{S}(f)$ is a term that models collisions and/or ionization. When $\overline{S} = 0$, the above equation is called *Vlasov equation*. The fluid models discussed in the previous paragraphs can be viewed as approximations to the Vlasov (or Boltzmann) equation obtained by taking moments in velocity space; i.e. by multiplying Eq. (20) by $v^n$, with $n$ = 0, 1, 2… and integrating over all velocities. The ion density, average velocity and kinetic temperature are defined respectively as the zeroth, first and second moment of the distribution function:

$$\begin{aligned} n_i &= \int f \, dv, \\ u_i &= \frac{1}{n_i} \int f \, v \, dv, \\ k_B T_i &= \frac{1}{n_i} \int f \, (v - u_i)^2 \, dv. \end{aligned} \qquad (21)$$

In the DS, Eq. (20) must be coupled to the Poisson equation (3) for the electrostatic potential. Using this system, a kinetic version of Bohm's criterion can be obtained, which reads [3, 15]

$$\left\langle v^{-2} \right\rangle \leq \frac{m_i}{k_B T_e}, \qquad (22)$$

where the average is computed using the ion distribution $f$, as in Eqs. (21). When the ions have zero temperature, Eq. (22) reduces to the usual Bohm criterion (7). Notice that the left-hand side Eq. (22) gives a particularly large weight to slow particles. Therefore, the ion distribution must decrease fast enough for $v \to 0$, so as to avoid the possible singularity. In practice, however, the features of the kinetic DS do not differ qualitatively from those obtained by means of the fluid analysis presented in the previous sections.

In order to model the presheath, the source term $\overline{S}$ must be specified, and Boltzmann's equation must be supplemented by the quasineutrality relation (for the presheath alone) or the full Poisson equation (to describe both the DS and the presheath). Notice that, for a purely collisional model (no ionization), the source term must satisfy $\int \overline{S} \, dv = 0$, so that the continuity equation (1) is unaffected. On the other hand, a collisionless model with ionization must satisfy $\int \overline{S} v \, dv = 0$, so that the momentum equation (2) is unaffected.

Let us now consider the Boltzmann equation (20) at steady state. The purely ionizing, collisionless case was first studied by Tonks and Langmuir [1] in a paper that set the beginnings of all further works on plasma–wall interactions. They selected a source of cold ions, $\bar{S} = S(x)\delta(v)$, where $\delta$ is the Dirac function and $S(x)$ represents the number of ions created per unit volume per second, which is usually taken as in Eq. (19). As there are no collisions, ions generated at a rate $S(z)dz$ in the volume element $dz$ attain the point $x$ at a velocity

$$v = \sqrt{2e\frac{\Phi(z) - \Phi(x)}{m_i}} \,. \tag{23}$$

At steady state, the elementary flux $d\Gamma$ at position $x$ is equal the number of ions created in the volume $dz$. Therefore

$$d\Gamma \equiv f(x,v)\, v\, dv = S(z)\, dz \,. \tag{24}$$

Dividing by $v$, integrating over velocity space and using Eq. (23), yields the ion density. When this is substituted into Poisson's equation (3) together with Boltzmann's relation for the electrons, one obtains

$$\frac{d^2\Phi}{dx^2} = -\frac{e}{\varepsilon_0}\left\{\left(\frac{m_i}{2e}\right)^{1/2}\int_{-\infty}^{x}\frac{S(z)\,dz}{\sqrt{\Phi(z) - \Phi(x)}} - n_0\exp\left(\frac{e\Phi(x)}{k_B T_e}\right)\right\}. \tag{25}$$

This is the so-called *plasma–sheath equation* (in one-dimensional planar geometry), and it describes both the presheath and the DS. In the limit of zero Debye length, the left-hand side Eq. (25) can be neglected and we obtain

$$\left(\frac{m_i}{2e}\right)^{1/2}\int_{-\infty}^{x}\frac{S(z)\,dz}{\sqrt{\Phi(z) - \Phi(x)}} = n_0\exp\left(\frac{e\Phi(x)}{k_B T_e}\right), \tag{26}$$

which describes the quasineutral presheath.

In their original paper, Tonks and Langmuir [1] solved Eq. (26) for the source function given in Eq. (19), with $\gamma = 0$ and 1, for planar, cylindrical and spherical geometry. Their treatment relies on a series expansion for the potential, the coefficients of which are computed numerically.

Harrison and Thompson [15] obtained an analytical solution to Eq. (26) in closed form (for planar geometry), expressed in terms of special functions. Their solution shows that the electric field becomes infinite at a certain point, which is independent of the particular source function employed. The appearance of a singularity is an indication that, beyond this point, the model is no longer valid. Indeed, such a point can be identified with the entrance of the DS: as within the DS the plasma is nonneutral, obviously the quasineutral Eq. (26) fails to be correct there.

Notice that the two studies described above address the same physical issues as the paper by Kino and Shaw [6] described in Sec. II.2 (i.e. cold source, collisionless regime, and quasineutrality), but Kino and Shaw used a fluid model. Fluid equations are an approximation of a more exact kinetic model, and therefore some physics is lost: however, they are usually more easily amenable to analytical studies. Qualitatively, however, results obtained with fluid and kinetic models reproduce in a similar way the main features of the presheath. In particular, the singular point at the entrance of the DS is a physical characteristic revealed by both types of models [see Eq. (17) and related discussion in Sec. II.2].

The complete plasma–sheath equation was first solved numerically by Self [16] for several values of $\gamma$ and different geometries. The results (obtained for $0.001 \leq \lambda_D / L_c \leq 0.1$) show a smooth transition between the quasineutral presheath and the nonneutral DS.

The above treatment was generalized to the case of a warm source by Emmert et al. [17], who took

$$\bar{S}(x,v) = H(x) \frac{m_i |v|}{2k_B T_s} \exp\left(-\frac{m_i v^2}{2k_B T_s}\right), \qquad (27)$$

where $H(x)$ is the spatial variation of the source strength and $T_s$ the source 'temperature'. Equation (27) represents a constant flux of particles produced with a Maxwellian distribution, and is the necessary choice to maintain a Maxwellian distribution far from the source. In another work, Bissel and Johnson [18] used a similar model, with source function

$$\bar{S}(x,v) = H(x) \frac{m_i}{2k_B T_s} \exp\left(-\frac{m_i v^2}{2k_B T_s}\right), \qquad (28)$$

which represents, for example, the source arising from the velocity-independent ionization of a Maxwellian distribution of neutrals, and $H(x)$ is proportional to the electron density. Both models actually yield similar ion distribution functions (both not Maxwellian) at the DS edge and at the wall. Experiments performed by Pitts [19] using a retarding-field analyzer (RFA, see also Sec. III) were not able to discriminate between the two models. The model of Emmert et al. was also used to interpret measurements of the ion distribution function obtained by laser induced fluorescence in an argon plasma [20].

The kinetic theory of a *collision-dominated* plasma–wall transition (no ionization) was studied by Riemann [21, 22]. In this case, the plasma is dominated by charge exchange collisions, which result simply in an interchange of the ion and neutral velocities. By assuming that the neutrals are cold, we obtain the following source term [see Eq. (20)]

$$\bar{S}(x,v) = -\frac{v}{\lambda(v)} f(x,v) + \delta(v) \int_{-\infty}^{\infty} \frac{w}{\lambda(w)} f(x,w) \, dw, \qquad (29)$$

where $\lambda(v)$ representing the (possibly velocity-dependent) mean free path. Notice that the integral of this term in velocity space gives no contribution, so that it indeed represents a purely collisional situation. Results obtained in the quasineutral limit indicate that the ion distribution function is approximately half-Maxwellian in the plasma core, but is strongly deformed when approaching the sheath. At the sheath edge, the kinetic Bohm criterion, Eq. (22), is satisfied marginally, whereas the original Bohm criterion is oversatisfied (i.e. $u_i > c_s$).

In summary, the plasma–wall transition is characterized by at least three different length scales, namely the Debye length $\lambda_D$, the collisional/ionizing length $L_c$, and the ion Larmor radius $\rho_i$. These correspond respectively to the typical extension of the Debye sheath (nonneutral), the collisional/ionizing presheath and the magnetic presheath (both quasi-neutral). The Debye length is generally much smaller than the other two lengths, whereas $L_c$ and $\rho_i$ can be of the same order of magnitude. For weakly collisional plasmas, it is reasonable to assume the ordering $\lambda_D \ll \rho_i < L_c$.

## III. INFLUENCE OF THE SHEATH ON PROBE MEASUREMENTS

Like every solid surface in contact with a plasma, probes are subject to the formation of a sheath and a presheath in their vicinity. These transition layers modify the plasma characteristics, so that the quantities measured by the probe do not actually correspond to the ones in the plasma. This is a general fact, not restricted to the case of plasmas and encountered for various kinds of experiments: probes influence the medium they measure. A good understanding of this interaction is therefore required to deduce correctly the values that would be measured without the probe perturbation.

The study of plasma–probe interaction is illustrated here for measurements of the ion temperature. Its determination is a difficult task, especially for fusion plasmas, since hydrogen isotopes are fully ionized and do not emit photons [23]. Thus, a spectroscopy diagnostic cannot provide directly the ion temperature, which can only be estimated from that of the neutrals [24-26]. As the coupling between ions and neutrals is not clearly understood, this technique is not very accurate. Moreover, it necessarily needs to average the temperature over some volume of plasma. An alternative technique, using a RFA device (retarding field analyzer) [27-30] that measures directly the ion energy distribution, is investigated here by means of a kinetic model. The equations are solved numerically and the results are interpreted in light of simplified analytical solutions.

The experimental device RFA has already been employed in various domains of plasma physics [31-34]. However, due to the high heat fluxes and the smallness of the Debye length of fusion plasmas, it has had a limited use in tokamaks (toroidal fusion devices confining the plasma by means of a strong magnetic field) [19, 27-29]. Nowadays, even if those problems can be overcome by an appropriate design of the analyzer [30, 35], there are still some difficulties. In particular, the area probed by the RFA corresponds to the boundary plasma, called the *scrape-off layer* or SOL (located in the vicinity of the material components of the tokamak chamber), where large flows exist [36-39]. These flows can affect significantly the ion current collected by the RFA, and consequently also the ion temperature estimation.

### III.1 Modeling of a RFA in a strong magnetic field

A schematic description of the analyzer is represented in Fig. 4a. It consists of a small entrance slit in the probe surface, two grids and a collector. The probe is aligned along the magnetic field lines so as to measure the parallel component of the ion flux. The entrance slit is sufficiently biased to a negative (and constant) potential, so that most of the electrons coming from the plasma are repelled. The retarding potential $\Phi_S$ applied to the first grid ranges from zero to large positive values in order to scan the ion distribution function. Only the ions with a kinetic energy larger than $e\Phi_S$ are collected. The second grid is negatively biased to a negative (constant) voltage to repel energetic electrons from the plasma (which pass the barrier potential of the entrance slit) and cancel out secondary electron emission created by ion impact on the collector. The entrance slit width is of the order of a Debye length or less [30], so that it is shielded by the sheath. In this case, the ion distribution function entering the analyzer is reasonably close to the one reaching its external surface, and most incident electrons are repelled back into the plasma.

In order to model the plasma–probe interaction in the SOL, following the approach of Chung and Hutchinson [40], we consider a fixed wall in contact with a collisionless, strongly magnetized, flowing plasma. The probe surface is perpendicular to the uniform magnetic field, so that no magnetic presheath is present. Therefore, the density perturbation caused by the probe is characterized by two regions: a Debye sheath and a quasineutral presheath. In typical tokamak edge plasmas, the Debye sheath thickness is of the order of 0.1 mm, whereas the diameter of the probe tends to be a few centimeters. On a macroscopic scale the Debye sheath is therefore negligible. The quasineutral presheath region extends along the field lines inside the flux tube connected to the probe. The presheath length is determined by the balance between the parallel flow normal to the probe surface and the cross-field transport that feeds the presheath from the unperturbed plasma outside the flux tube (Fig. 4b).

The probe considered here is a double-mounted RFA, which can provide simultaneous measurements from both sides. In most tokamak SOLs, with magnetic field strengths of a few Teslas and ion temperatures some tens of electron volts, the Larmor radius is typically a few tenths of a millimeter and is consequently much smaller than the size of the probe. In this case, the cross-field transport can be considered as being diffusive [40] and is modeled as a random migration of ions across magnetic field lines. The migration rate is governed by the magnetic field strength, so that, for typical SOL regimes, parallel convection dominates over perpendicular transport. Therefore, the parallel length of the presheath $L_{//}$ is very long compared to the cross-field dimension of the probe $L_\perp$ [41]. The study can therefore be reduced to a one-dimensional model in the parallel direction on condition that the cross-field transport is included. As introduced in Sec. II.4, the relevant equation for weakly collisional plasmas is the Vlasov equation (20). In this case, $\bar{S}$ does not model collisions nor ionization, but the migration of ions across magnetic field lines. We assume that this migration occurs, in both directions, at a constant frequency $W = D_\perp / L_\perp^2$, where $D_\perp$ is the cross-field diffusion coefficient [40]. Then, $\bar{S}$ takes the form

$$\bar{S} = W(f_0 - f) \tag{30}$$

where $f$ and $f_0$ are respectively the ion distribution function in the presheath and in the unperturbed plasma. The first term on the right hand side of Eq. (30) models ions entering the presheath from the unperturbed plasma and the second ions exchanged in the other direction. Assuming electrons at thermal equilibrium, their density in the presheath is given by the Boltzmann relation (4). The set of equations (4), (20), and (30) is closed self-consistently by using the quasineutrality condition ($n_i = n_e$) [Eq. (16)]. In order to express positions and velocities with appropriate units, we use the following transformations

$$t \to W t, \quad u \to \frac{v}{c_s}, \quad x \to \frac{Wx}{c_s}, \quad \text{and} \quad \Phi \to \frac{e\Phi}{k_B T_e}, \tag{31}$$

where $c_s$ is the acoustic speed given by Eq. (7). Equations (16), (20), (21), and (30) can thus be written in the compact form

$$\frac{\partial f}{\partial t} + u\frac{\partial f}{\partial x} - \frac{e}{m_i}\frac{\partial \Phi}{\partial x}\frac{\partial f}{\partial u} = f_0 - f \tag{32}$$

$$\Phi(x,t) = \ln \int f \, du . \tag{33}$$

Outside the presheath, the ion distribution function $f_0$ is assumed to be a shifted Maxwellian with temperature $T_{i0}$ and mean velocity $U_0$. The expression of $f_0$ is thus

$$f_0 = \frac{1}{\sqrt{2\pi\tau}} \exp\left(-\frac{(u-U_0)^2}{2\tau}\right), \tag{34}$$

where $\tau = T_{i0} / T_e$.

We assume that the probe surface (located at $x = 0$) is perfectly absorbing, and that far from the probe the ion distribution is equal to $f_0$. Therefore, using Eqs. (33)-(34), the boundary conditions become

$$f(x = 0, u > 0) = 0, \quad f(x \to \infty) = f_0, \quad \Phi(x \to \infty) = 0 \tag{35}$$

on the upstream side and

$$f(x = 0, u < 0) = 0, \quad f(x \to -\infty) = f_0, \quad \Phi(x \to -\infty) = 0 \tag{36}$$

on the downstream side (see Fig. 4b). The upstream (downstream) side is defined as the side of the probe where the mean velocity $U_0$ is directed towards (opposite to) the probe surface.

Therefore, considering this set of equations, the presheath behavior is governed by only two dimensionless parameters of the unperturbed plasma, namely the ion-to-electron temperature ratio $\tau$ and the mean velocity $U_0$ (normalized to $c_s$), also called the plasma drift velocity. As we are dealing with a quasineutral regime, the DS is outside the scope of the present analysis.

### III.2 Presheath structure

A complete resolution of Eqs. (32)-(33) requires a numerical treatment: to obtain the present results, we used a Vlasov-Eulerian code [42-44], which computes self-consistently the ion distribution function $f$ and the electric potential $\Phi$ until a stationary equilibrium is reached.

Nevertheless, by neglecting the electric field ($\partial \Phi / \partial x = 0$) and at equilibrium ($\partial . / \partial t = 0$), it is possible to solve analytically Eqs. (32)-(33). With the boundary conditions specified in Eqs. (35)-(36), a solution for $f$ is

$$f(x,u) = f_0(u)\left(1 - H(\pm u)\,\exp\left(-\frac{x}{u}\right)\right), \tag{37}$$

where positive and negative signs stand respectively for upstream and downstream cases, and *H* is the Heaviside function. It will be useful to contrast this analytical solution with the numerical results in order to distinguish the importance of simple geometrical effects with respect to distortion of the distribution function by the electric field.

For the purpose of illustration, the two parameters controlling the presheath behavior are set to $\tau = 2$ and $U_0 = 1$. This is in agreement with expectations in the SOL plasmas [36-39], where ions are generally warmer than electrons and where large flows are present. Figure 5a-b presents the ion distribution function $f(x,v)$ computed numerically, on both sides of the analyzer, at different positions. It shows the progressive depletion of the distribution from the plasma boundary to the probe surface, where particles are absorbed [cf. Eqs. (35)-(36)]. Furthermore, in the vicinity of the probe, the distributions are clearly not Maxwellian, particularly on the downstream side, which points out that a kinetic model is indeed necessary. Figure 6 shows the electrostatic potential and ion kinetic temperature, the latter being defined in Eq. (21). Profiles are characterized by steeps gradients near $x = 0$, showing that an isothermal assumption for the ions would not be appropriate.

Since the plasma is flowing at the drift velocity $U_0$, Fig. 4 is asymmetric, so that on each side of the probe the plasma perturbations are not the same. The probe shadows the ions coming from the upstream direction (with positive velocities) on the downstream side, and vice versa on the upstream side. This shadowing effect yields a density decrease and thus a potential drop, which are both larger on the downstream side. Figures 5-6 show quantitatively these differences and indicate that the presheath structure depends strongly on $U_0$. Considering the drastic assumption made on the electric field, the analytical profiles are rather close to those obtained by solving numerically the full Vlasov equation. This shows that the electric field is not predominant: the presheath behavior is mainly governed by the probe shadowing.

### III.3 Ion temperature estimation

The retarding potential $\Phi_S$ applied to the ion selector of the RFA (see Fig. 4a) affects only the ions whose kinetic energy is too small to reach the collector. Therefore, the differential fluxes $u f_i(x=0, u)$ at the wall can be scanned by varying $\Phi_S$ on each side. Provided that the potential on the collector is lower than the one at the entrance slit, all the ions passing through the ion selector reach the collector. The downstream and upstream collected currents are then

$$J_{C,Down} = \int_{\sqrt{2\Phi_s}}^{\infty} u f(x=0,u)\,du, \qquad J_{C,Up} = \int_{-\infty}^{-\sqrt{2\Phi_s}} u f(x=0,u)\,du. \tag{38}$$

Collected currents vary from their maximum values (obtained for $\Phi_S = 0$) to zero (for large positive values of $\Phi_S$). RFA characteristics, which are the semi-logarithm plots presented in Fig. 7, are the kind of results obtained experimentally. According to Pitts [30], an estimated ion temperature $T_{RFA} = -1/\alpha$ can be deduced from the slope $\alpha$ of the linear part of the characteristics. However, this estimation would give the correct ion temperature only if the ion distribution on the wall were a half-Maxwellian with no shift, which is clearly not the case, as shown in Fig. 5a-b. The kinetic modifications can be significant, so it is not surprising that the measured value is not equal to the equilibrium plasma temperature $T_{i0}$ ($\tau$ in normalized unit). On the downstream side, the analytical curve in Fig. 7 does not fit well the numerical data. However, the temperature estimation remains close to the numerical one, as only the slope of its linear part matters. Therefore, both numerical and analytical approaches can be used to investigate the relationship between the measured temperature $T_{RFA}$ and the plasma temperature $\tau$. Figure 8 shows the dependence of the measurements on $U_0$. $T_{RFA}$ is always overestimated on the upstream side, and underestimated on the downstream side; $T_{RFA}$ coincides with $\tau$ only for $U_0 = 0$. Indeed, measurements on both sides counterbalance each other [45], so that the average RFA temperature defined as

$$\bar{T}_{RFA} \equiv \frac{T_{RFA,up} + T_{RFA,down}}{2} \tag{39}$$

gives a value close to $\tau$. It can be shown analytically that $\bar{T}_{RFA}$ is equal to $\tau$ to second order in a power expansion with respect to $U_0$, whereas $T_{RFA}$ is equal to $\tau$ only to zeroth order. Table 1 shows the temperatures measured on each side of the probe and the averaged values, for a plasma drifting at the acoustic speed ($U_0 = 1$). Single measurements on either side of the probe lead to an estimation of $\tau$ with an error up to 40 %, whereas the relative difference between $\bar{T}_{RFA}$ and $\tau$ is only about 5 %.

| $U_0 = 1$ | $T_{RFA,down}$ | $T_{RFA,up}$ | $\bar{T}_{RFA}$ |
|---|---|---|---|
| $\tau = 1$ | 0.66 (0.66) | 1.44 (1.46) | 1.05 (1.06) |
| $\tau = 2$ | 1.51 (1.50) | 2.58 (2.62) | 2.05 (2.06) |
| $\tau = 3.5$ | 2.84 (2.81) | 4.22 (4.30) | 3.53 (3.56) |
| $\tau = 5$ | 4.20 (4.16) | 5.88 (5.94) | 5.04 (5.05) |

*Table 1: Numerically-computed downstream, upstream and average RFA temperatures (normalized to $T_e$) for $U_0 = 1$. Analytical results are in parentheses.*

Therefore, even if the RFA temperature on each side of the analyzer is not equal to the one in the plasma, an accurate estimation can be obtained from Eq. (39). Furthermore, as $\bar{T}_{RFA}$ does not depend on $U_0$, it is not necessary to know its exact value. This is an important point for tokamak physics, as the drift velocity $U_0$ is often difficult to assess by direct measurement.

In summary, this section clearly shows that great care should be used when interpreting experimental data obtained from probes. Indeed, due to the presence of the sheath and the presheath in the vicinity of the probe, the measured value of a quantity may differ significantly from the 'real' one in the plasma core. Only a careful analysis of the plasma–wall transition can enable us to relate the former to the latter.

## IV. CONCLUSION

In this paper, we have reviewed some basic concepts of plasma–wall interaction, focussing in particular on the physics of sheath formation. This is an important topic in many areas of plasma and nuclear fusion research. Naturally, the geometric configurations pertinent to magnetic fusion devices are almost invariably more complex than the one-dimensional models adopted here. However, such simplified pictures are often surprisingly accurate in reproducing the main features of the plasma–wall transition. As in many areas of plasma physics, models of two types are available: fluid and kinetic. The former are more easily solved numerically, and sometimes amenable to analytical treatment; however, as it was seen in Sec. III, the distribution functions in the (pre)sheaths are generally far from being Maxwellian, so that a full kinetic model is often indispensable.

All models indicate that the plasma–wall transition is composed of several regions: the electrically charged Debye sheath just in front of the wall, followed by several presheaths, where collisions, ionization or magnetic effects dominate and the plasma is quasineutral. The global structure of the plasma–wall transition is determined by the relative importance of such effects. In most applications, the DS is much thinner than the other regions, whereas the various presheaths can be of comparable thickness, depending on the physical regime of interest.

The structure of the sheaths is of great importance for magnetic fusion devices, because it determines the way particles (and energy) are collected at the external walls of the tokamak and eventually recycled. Power deposition on the walls is indeed a crucial parameter to ensure a self-sustaining, steady-state operation of a fusion reactor. Besides, sheath formation affects (sometimes in a dramatic way) measurements obtained with probes, because the sheath can alter the value of the quantity that is being measured. A recent application to retarding field analyzers was presented in Sec. III — the results indeed showed that, due to the sheath influence, the ion temperature measured on one side of the analyzer can differ significantly from that of the unperturbed plasma. However, by measuring the temperature on both sides of the RFA and averaging the results, one can obtain a much more accurate estimation. This is a nice example of how a detailed theoretical analysis of the sheath structure can lead to very practical implications for probe measurements.

**Acknowledgments**

We are indebted to Dr. J.P. Gunn for his expertise on RFA probes and for the fruitful collaboration that led to the results reported in Sec. III.

# References

[1] L. Tonks and I. Langmuir, Phys. Rev. **34**, 876 (1929).
[2] P.C. Stangeby, *The plasma boundary of magnetic fusion devices* (Institute of Physics Publishing, London, 2000).
[3] K.-U. Riemann, J. Appl. Phys. D **24**, 493 (1991).
[4] F.F. Chen, *Introduction to plasma physics and controlled fusion* (Plenum Press, New York, 1984).
[5] D. Bohm, *The characteristics of electrical discharges in magnetic field*, ed. A. Guthry and R.K. Wakerling (MacGraw-Hill, New York, 1949.
[6] G. Kino and E.W. Shaw, Phys. Fluids **9**, 587 (1966).
[7] R.N. Franklin and J. Snell, Phys. Plasmas **7**, 3077 (2000).
[8] R. Chodura, Phys. Fluids **25**, 1628 (1982).
[9] F. Valsaque and G. Manfredi, J. Nucl. Mater. **290-293**, 763 (2001).
[10] K.-U. Riemann, Phys. Plasmas **1**, 552 (1994).
[11] K.-U. Riemann, Contrib. Plasma Phys. **34**, 127 (1994).
[12] P.C. Stangeby, Phys. Plasmas **2**, 702 (1995).
[13] H. Schmitz, K.-U. Riemann, and Th. Daube, Phys. Plasmas **3**, 2486 (1996).
[14] Th. Daube and K.-U. Riemann, Phys. Plasmas **6**, 2409 (1999).
[15] E.R. Harrison and W.B. Thompson, Proc. Phys. Soc. **74**, 145 (1959).
[16] S. A. Self, Phys. Fluids **6**, 1762 (1963).
[17] G.A. Emmert, R.M. Wieland, A.T. Mense, and J.N. Davidson, Phys. Fluids **23**, 803 (1980).
[18] R.C. Bissel and P.C. Johnson, Phys. Fluids **30**, 779 (1987).
[19] R.A. Pitts, Phys. Fluids B **3**, 2871 (1991).
[20] G. Bachet, L. Chérigier, and F. Doveil, Phys. Plasmas **2**, 1782 (1995).
[21] K.-U. Riemann, Phys. Fluids **24**, 2163 (1981).
[22] K.-U. Riemann, J. Appl. Phys. D **25**, 1432 (1992).
[23] G.F. Matthews, R. A. Pitts, G.M. McCracken, and P.C. Stangeby, Nucl. Fusion **31**, 1495 (1991).
[24] H. Weisen, M. Von Hellermann, A. Boileau, L.D. Horton, W. Mandl, and H.P. Summers, Nucl. Fusion **29**, 2187 (1989).
[25] M. Von Hellermann, W. Mandl, H.P. Summers, H. Weisen, A. Boileau, P.D. Morgan, H. Morsi, R. Koenig, M.F. Stamp, and R. Wolf, Rev. Sci. Instrum. **61**, 3479 (1990).
[26] M.F. Stamp and H.P. Summers, *Proceedings of Contributed Papers*, 17th European Conference on Controlled Fusion and Plasma Heating, Amsterdam, European Physical Society 1377 (1990).
[27] G.F. Matthews, J. Phys. D **17**, 2243 (1984).
[28] A.S. Wan, T.F. Yang, B. Lipschultz, and B. LaBombard, Rev. Sci. Instrum. **57**, 1542 (1986).
[29] R.A. Pitts, Ion Energy, *Sheath Potential and Secondary Electron Emission in the Tokamak Edge*, PhD Thesis, University of London, 1991.
[30] R.A. Pitts, Contrib. Plasma Phys. **36**, 87 (1996).
[31] C.P. DeNeef and A.J. Theiss, Rev. Sci. Instrum. **50**, 378 (1979).
[32] A.W. Molvik, Rev. Sci. Instrum. **52**, 704 (1981).
[33] S.G. Ingram and N. St. J.Braithwaite, J. Appl. Phys. **68**, 5519 (1990).
[34] C. Böhm and J. Perrin, Rev. Sci. Instrum. **64**, 31 (1993).
[35] G.F. Matthews, Plasma Phys. Control. Fusion **36**, 1595 (1994).
[36] C. Boucher, C.S. MacLatchy, G. Le Clair, J.L. Lachambre, and M. St-Onge, J. Nucl. Mater. **176-177**, 1050 (1990).

[37] N. Tsois, C. Dorn, G. Kyriakakis, M. Markoulaki, M. Pflug, G. Schramm, P. Theodoropoulos, P. Xantopoulos, M. Weinlich, et al., J. Nucl. Mater. **266-269**, 1230 (1999).

[38] A. Hatayama, H. Segawa, R. Schneider, D.P. Coster, N. Hayashi, S. Sakurai, N. Asakura, and M. Ogasawara, Nucl. Fusion **40**, 2009 (2000).

[39] S.K. Erents, A.V. Chankin, G.F. Matthews, and P.C. Stangeby, Plasma Phys. Control. Fusion **42**, 905 (2000).

[40] K-S. Chung and I.H. Hutchinson, Phys. Rev. A **38**, 4721 (1988).

[41] F. Valsaque and G. Manfredi, *Proceedings of Contributed Papers*, 28th European Conference on Controlled Fusion and Plasma Physics, Funchal, European Physical Society, to be published (2001).

[42] C.Z. Cheng and G. Knorr, J. Computational Phys. **22**, 330 (1976).

[43] G. Manfredi, M. Shoucri, R. O. Dendy, A. Ghizzo, and P. Bertrand, Phys. Plasmas **3**, 202 (1996).

[44] E. Sonnendrücker, J. Roche, P. Bertrand, and A. Ghizzo, J. Computational Phys. **149**, 201 (1999).

[45] F. Valsaque, G. Manfredi, J. P. Gunn, and E. Gauthier, submitted to Phys. Plasmas.

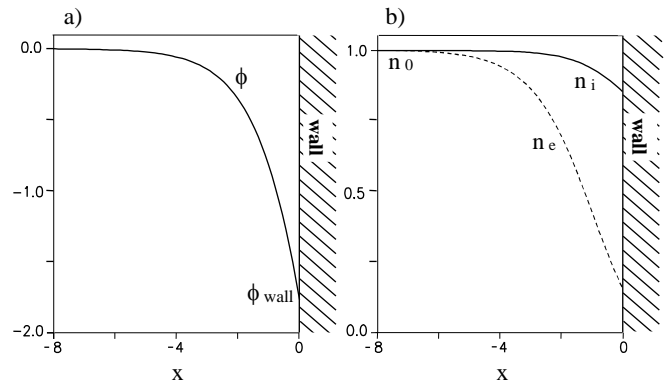

FIG. 1: Typical structure of the Debye sheath. a) Electrostatic potential (normalized to $k_B T_e/e$); b) electron and ion densities (normalized to the plasma density $n_0$).

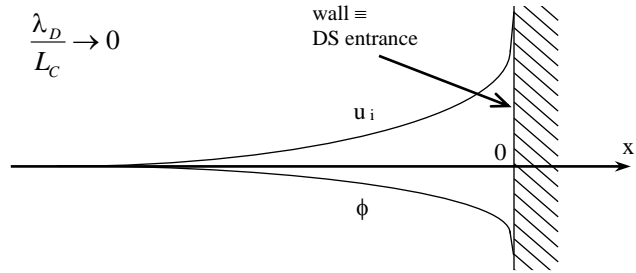

FIG. 2: Schematic behavior of the mean ion velocity $u_i$ and electric potential $\Phi$ on the presheath scale ($\lambda_D / L_C \to 0$). Notice the singularity (infinite derivative) at the DS entrance.

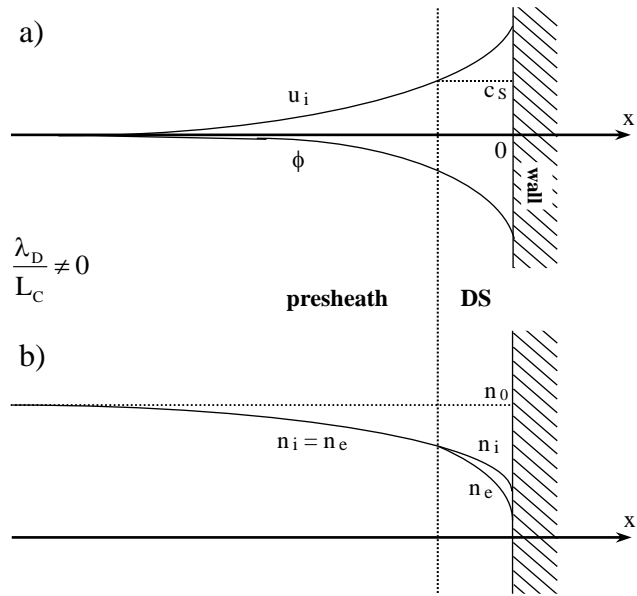

FIG. 3: Schematic description of the presheath and the DS when $\lambda_D / L_C$ is small but finite. a) Mean ion velocity and electric potential; b) ion and electron densities.

FIG. 4: a) Experimental device: schematic view of the Retarding Field Analyzer, on the upstream side. b) Double RFA in a flowing plasma, in the presence of a strong magnetic field.

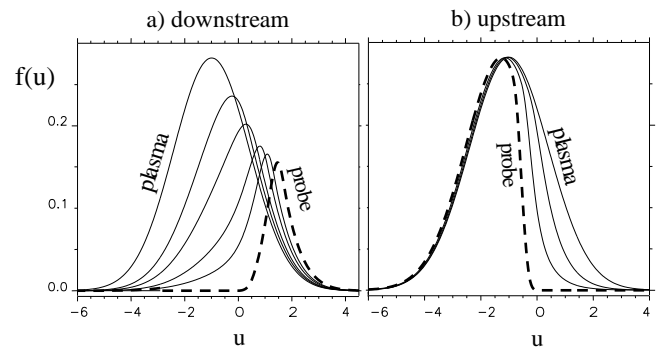

FIG.5: Numerical ion distributions ($\tau = 2$ and $U_0 = 1$), for different positions from the plasma to the probe surface (dashed curve). a) Downstream side; b) upstream side.

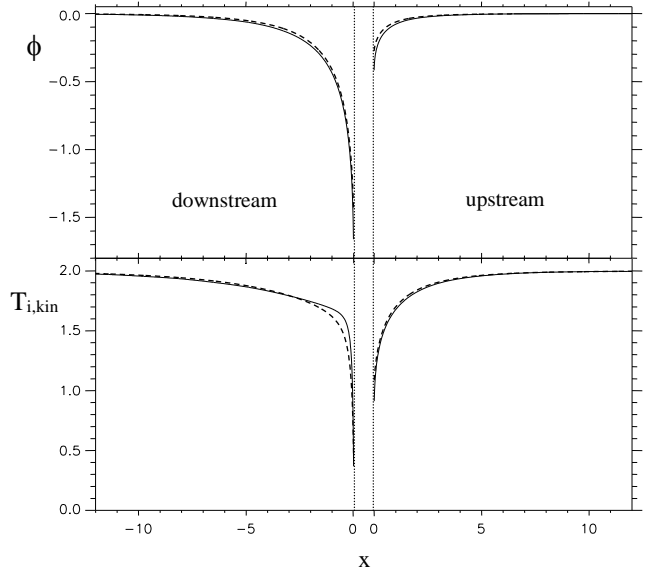

FIG. 6: Electric potential and ion kinetic temperature, respectively normalized to $k_B T_e/e$ and $T_e$, versus position normalized to $c_S/W$ (numerical results: solid line; analytical model: dashed line), on each side of the RFA (for $\tau = 2$ and $U_0 = 1$).

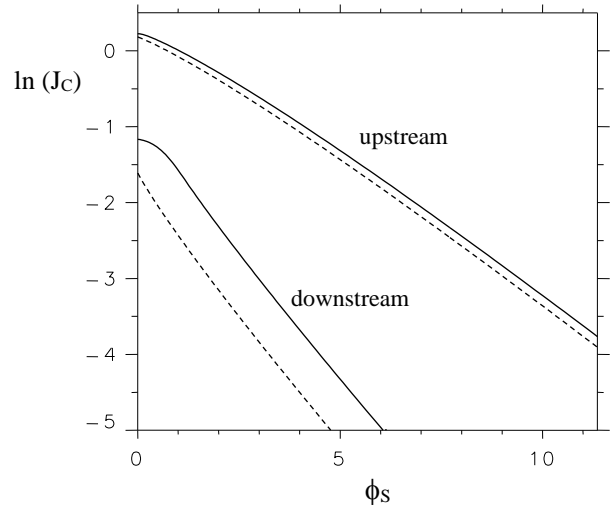

FIG. 7: Numerical (solid line) and analytical (dotted line) RFA characteristics, for $\tau = 2$ and $U_0 = 1$. (Current and potential respectively normalized to $n_0 c_s$ and $k_B T_e/e$.)

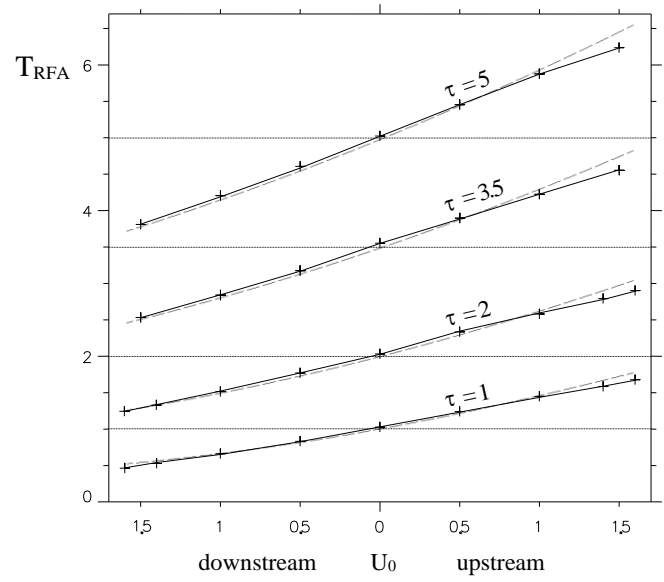

FIG. 8: Numerical (solid line) and analytical (dotted line) RFA temperatures (normalized to $T_e$), for $\tau = 1, 2, 3.5$, and $5$.